\documentclass[letterpaper,10pt]{article} 
\usepackage{opticameet3} %% use version 3 for proper copyright statement

\newcommand\authormark[1]{\textsuperscript{#1}}

\usepackage{amsmath,amssymb}
\usepackage{array} % for the table
\usepackage[colorlinks=true,bookmarks=false,citecolor=blue,urlcolor=blue]{hyperref} %pdflatex
\begin{document}

\title{Theoretical efficiency limit of diffractive input couplers in augmented reality waveguides}

\author{Zhexin Zhao,\authormark{1*} Yun-Han Lee,\authormark{1} Xiayu Feng,\authormark{1} Michael J Escuti,\authormark{1} Lu Lu,\authormark{1} and Barry Silverstein\authormark{1}}

\address{\authormark{1}Meta Reality Labs, Redmond, WA 98052, USA}

\email{\authormark{*}zhexinzhao@meta.com, zhexin@alumni.stanford.edu} %% email address is required; see note below about the corresponding author designation

\begin{abstract*} 
Considerable efforts have been devoted into augmented reality (AR) displays to enable the immersive user experience in the wearable glasses form factor. Transparent waveguide combiners offer a compact solution to guide light from the microdisplay to the front of eyes while maintaining the see-through optical path to view the real world simultaneously. To deliver a realistic virtual image with low power consumption, the waveguide combiners need to have high efficiency and good image quality. One important limiting factor for the efficiency of diffractive waveguide combiners is the out-coupling problem in the input couplers, where the guided light interacts with the input gratings again and get partially out-coupled. In this study, we introduce a theoretical model to deterministically find the upper bound of the input efficiency of a uniform input grating. Our model considers the polarization management at the input coupler and can work for arbitrary input polarization state ensemble. Our model also provides the corresponding characteristics of the input coupler, such as the grating diffraction efficiencies and the Jones matrix of the polarization management components, to achieve the optimal input efficiency. Equipped with this theoretical model, we investigate how the upper bound of input efficiency varies with geometric parameters including the waveguide thickness, the projector pupil size, and the projector pupil relief distance. Our study shines light on the fundamental efficiency limits on input couplers in diffractive waveguide combiners and highlights the benefits of polarization control in improving the input efficiency.
\end{abstract*}

%%%%%%%%%%%%%%%%%%%%%%%%%%  body  %%%%%%%%%%%%%%%%%%%%%%%%%%
\section{Introduction}
Augmented reality (AR) has attracted significant research attention from both academia and industry. AR technologies enable users to experience virtual contents in addition to the normal vision of the real world. 
From the early introduction of the AR concept \cite{sutherland1968head}, researchers have explored various AR architectures to pursue better display quality, smaller form factor, and wider accessibility \cite{kress2020optical, xiong2021augmented}. These AR architectures include free-space geometric combiners \cite{pulli201711, cheng2009design, wei2018design}, steered retinal projection \cite{lee2020foveated, kim2019foveated}, Maxwellian view displays \cite{maimone2017holographic, chang2020toward, jang2017retinal, kim2018optical}, and waveguide combiners \cite{yoshida2018plastic, kress2020waveguide, ayras2009exit, waldern2018digilens}. Among these architectures, waveguide combiners have a few advantages. (1) The waveguide combiners are compact. The waveguide combiner architecture typically contains a sub-millimeter-thick slab waveguide and a projector, e.g., liquid crystal on silicon (LCoS), micro light emitting diode (mLED), laser beam scanning (LBS) projector, or digital light projector (DLP). With the development of microdisplay, the projector can be integrated into the glasses frame. (2) The waveguide combiner can provide a large eyebox through pupil replication, where the etundue constraint is relaxed. On the contrary, in other architectures such as Maxwellian view displays, the etundue is conserved, such that it is challenging to achieve large field of view (FOV) and large eyebox simultaneously with a compact form factor. (3) Comparing with steered retinal projection, waveguide combiners do not require active tuning optical elements nor strongly depend on eye-tracking.  

\begin{figure}[tbp]
    \centering
    \includegraphics[width=9cm]{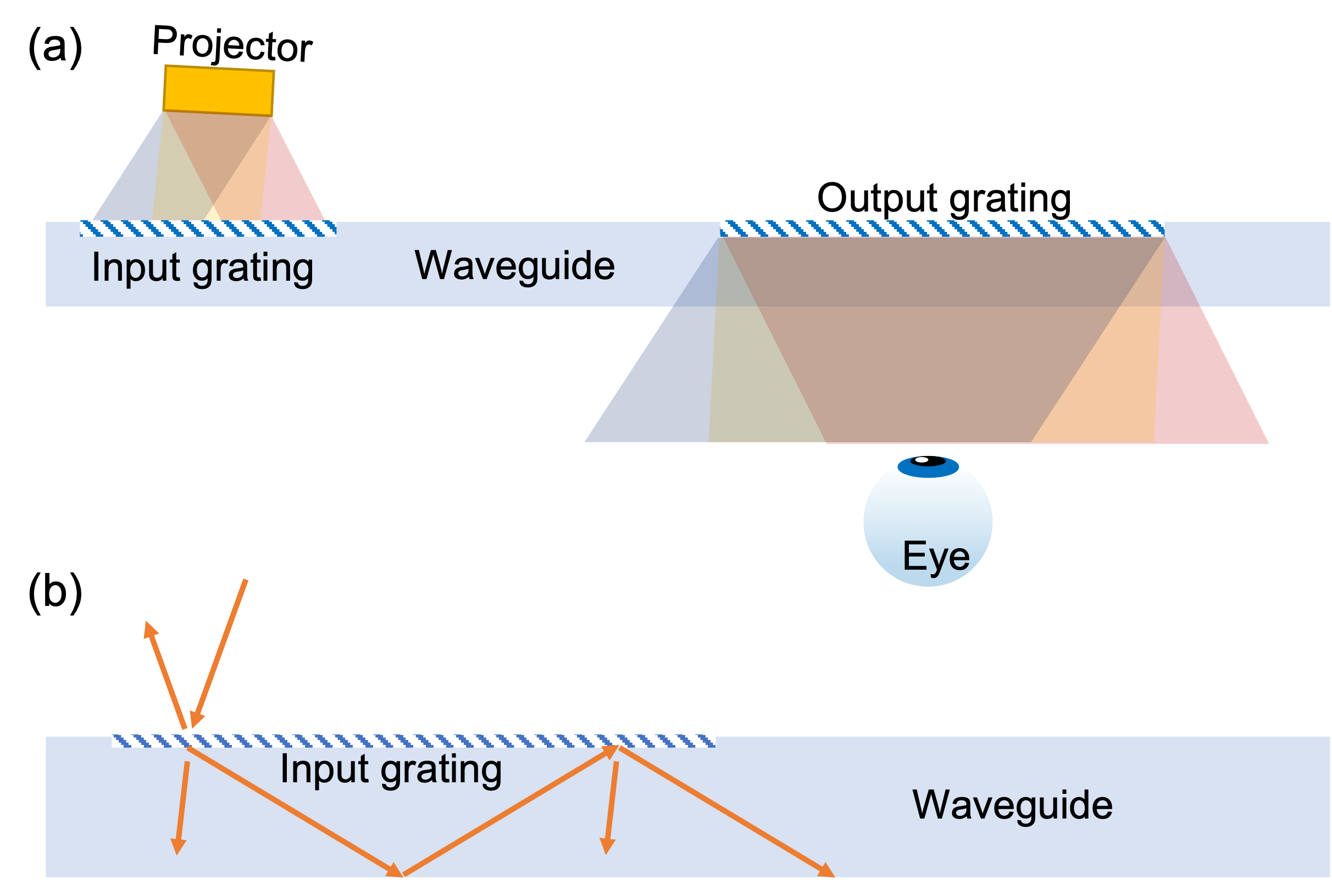}
    \caption{(a) A schematic of the diffractive waveguide combiner system, consisting of a projector, a dielectric slab waveguide with the input grating and the output grating. (b) An enlarged view of the input grating to illustrate the outcoupling problem in the input grating.}
    \label{fig:schematic}
\end{figure}

A waveguide combiner is typically made of transparent dielectric material with a thickness on the order of sub-millimeter. The input and output couplers are made of diffractive gratings, as in diffractive waveguide combiners, or semi-reflective mirrors, as in geometric waveguide combiners. In a diffractive waveguide combiner, collimated light from the projector is diffracted by the input grating and becomes guided in the waveguide. To confine light within the waveguide, the propagation angle of the guided light is larger than the critical angle for total internal reflection (TIR). The guided light then interacts with one output grating, in typical one-dimensional pupil expansions, or two gratings, in typical two-dimensional pupil expansions. After the interaction with the output grating, light propagates to the eye with the same direction as the light coming from the projector, as illustrated in Fig. \ref{fig:schematic}(a). 

With an extended etundue to achieve a large eyebox, the input power is spread into a larger spatial-angular phase space. Furthermore, part of the light cannot reach the eyebox. Therefore, brightness is the prominent bottleneck for waveguide combiners. In diffractive waveguide combiners, an important loss channel is at input couplers due to out-coupling. As illustrated in Fig. \ref{fig:schematic}(b), when the guided light interacts with the input grating, some of the guided light would be diffracted and outcoupled. Moreover, due to reciprocity, which will be discussed in detail later, the higher the diffraction efficiency of the grating, the larger the chance of getting out-coupled. To improve the efficiency of waveguide combiners, it is crucial to optimize the efficiency of input couplers and investigate the upper bounds set by physics. Towards this goal, Goodsell et al. studied the geometry-based efficiency limit for input couplers and designed metagratings to achieve the geometry-based efficiency limit \cite{goodsell2023metagrating}. Although their study explained the underline physics of the limited efficiency at input couplers and outlined a path to design metagratings to achieve the desired diffraction efficiencies, there are a few critical limitations of their work: (1) Their geometry-based efficiency limit works only when the projector pupil relief distance is zero. However, in many waveguide combiner systems, the projector pupil relief distance is larger than zero. (2) The polarization selectivity of the input grating is not considered. But, most gratings have different diffraction efficiency for each polarization. Furthermore, researchers have studied diffractive gratings with strong polarization selectivity and proposed polarization management to improve the performance of waveguide combiners \cite{lee2017reflective, xiong2021holographic, xiong2020rigorous, xiong2021planar, feng2021compensator,feng202130}. Thus, it is necessary to thoroughly investigate the efficiency limit of input couplers with general configurations and with the consideration of polarization management.

In this study, we theoretically analyze the input grating and study the upper bound of its efficiency. We provide numerical demonstrations on uniform input gratings, and we also generalize the theory to include spatial varying input gratings. Our theory works for polarization sensitive or insensitive input gratings and arbitrary incident polarization state ensemble. 
Our model can provide the upper bound of the input efficiency, along with the corresponding characteristics of the input coupler, such as the grating diffraction efficiencies and the Jones matrix of the polarization management components to achieve the optimal input efficiency. We theoretically demonstrate that polarization management can improve the upper bound of input efficiency. We also apply our theory to study how the optimal input efficiency of a uniform input grating varies with geometric parameters including the waveguide thickness, the projector pupil size, and the projector pupil relief distance. These discussions are crucial to design a highly efficient waveguide combiner system.
Furthermore, our study provides a solid theoretical foundation for approaches that utilizes polarization management to increase the input efficiency. 

\section{Theory}
\label{sec:theory}

We define the diffraction efficiency of the input grating as the ratio between diffracted power towards the target order and the incident power. In the contrast, we define the input efficiency as the ratio between the guided power in the waveguide after finishing the interactions with the input grating and the incident power from the projector. The diffraction efficiency and input efficiency are defined for each FOV.

\begin{figure}[tbp]
    \centering
    \includegraphics[width=7.5cm]{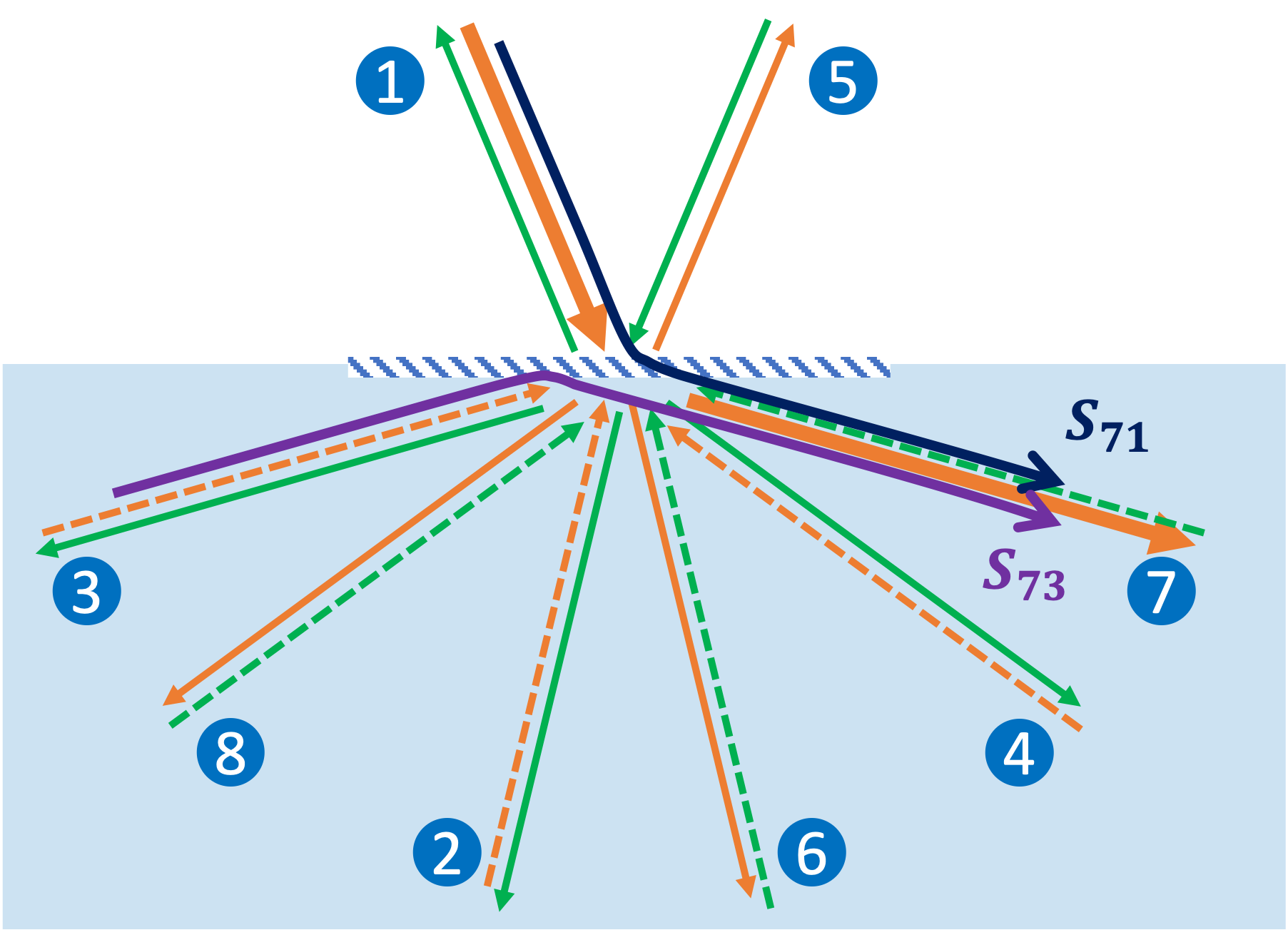}
    \caption{Illustration of the scattering channels of the input grating at the interface between air (top) and waveguide (bottom). The solid orange arrows indicate the possible power flow when light is incident from channel 1. The solid green arrows indicate the possible power flow when light is incident from channel 5. The dashed orange and green arrows represent all other possible incident or scattered power flow. The navy arrow highlights the diffraction from channel 1 to channel 7. The purple arrow highlights the reflection from channel 3 to channel 7.}
    \label{fig:scattering_channels}
\end{figure}

To calculate the input efficiency, we first investigate the scattering matrix of the input grating for each FOV. For a specific incident angle in air, we illustrate the scattering channels in Fig. \ref{fig:scattering_channels}, where we assume that only the zeroth diffraction order exists in the air, and three diffraction orders, -1-st, 0-th, and +1-st order, exist in the waveguide. For example, when light is incident from channel 1, channel 5 is the 0-th order reflection channel; channel 6 is the 0-th order transmission channel; and channels 7 and 8 are the +1-st and -1-st order diffraction channel respectively. 
With such channel labeling, the scattering matrix of this input grating is
\begin{equation}
    \label{eq:Smatrix}
    S = \begin{bmatrix}
    0 & 0 & 0 & 0 & S_{15} & S_{16} & S_{17} & S_{18} \\ 
    0 & 0 & 0 & 0 & S_{25} & S_{26} & S_{27} & S_{28} \\
    0 & 0 & 0 & 0 & S_{35} & S_{36} & S_{37} & S_{38} \\
    0 & 0 & 0 & 0 & S_{45} & S_{46} & S_{47} & S_{48} \\
    S_{51} & S_{52} & S_{53} & S_{54} & 0 & 0 & 0 & 0 \\
    S_{61} & S_{62} & S_{63} & S_{64} & 0 & 0 & 0 & 0 \\
    S_{71} & S_{72} & S_{73} & S_{74} & 0 & 0 & 0 & 0 \\
    S_{81} & S_{82} & S_{83} & S_{84} & 0 & 0 & 0 & 0
    \end{bmatrix}.
\end{equation}
Each block in this scattering matrix is a $2\times2$ matrix since there are two orthogonal polarizations. Typical input gratings satisfy Lorentz Reciprocity. Thus, the scattering matrix is symmetric, i.e., $S=S^T$ \cite{zhao2019connection, guo2022internal}. Furthermore, the input gratings are usually made of low-loss dielectric material and the absorption at the input gratings can be neglected. Under the assumption of energy conservation, the scattering matrix is unitary, i.e., $S^{\dagger} S=I$ \cite{zhao2019connection, guo2022internal}, where $I$ is an identity matrix. We also assume that higher order diffraction is negligible. For instance, when light incidents from channel 3, channel 8 is the -2-nd order diffraction, which is negligible. Moreover, many diffractive gratings in waveguide combiners have a slant feature, like slanted surface relief gratings \cite{bai2010optimization, miller1997design} or Bragg gratings with slanted Bragg plans \cite{waldern2018digilens, feng2021compensator, xiong2020rigorous, odinokov2020augmented, kress2019optical}, such that light is dominantly diffracted into one order.
we thus take an optimistic assumption that only one diffraction order dominates the diffraction for each incident channel and neglect the non-dominant diffraction order. These assumptions are also necessary to achieve the optimal input efficiency. Under these assumptions, we can simply the scattering matrix to
\begin{equation}
    \label{eq:Smatrix_simplified}
    S = \begin{bmatrix}
    0 & 0 & 0 & 0 & S_{51}^T & S_{61}^T & S_{71}^T & 0 \\ 
    0 & 0 & 0 & 0 & S_{52}^T & S_{62}^T & 0 & S_{82}^T \\
    0 & 0 & 0 & 0 & 0 & S_{63}^T & S_{73}^T & 0 \\
    0 & 0 & 0 & 0 & S_{54}^T & 0 & 0 & S_{84}^T \\
    S_{51} & S_{52} & 0 & S_{54} & 0 & 0 & 0 & 0 \\
    S_{61} & S_{62} & S_{63} & 0 & 0 & 0 & 0 & 0 \\
    S_{71} & 0 & S_{73} & 0 & 0 & 0 & 0 & 0 \\
    0 & S_{82} & 0 & S_{84} & 0 & 0 & 0 & 0
    \end{bmatrix}.
\end{equation}
Since the scattering matrix is unitary, we have
\begin{equation}
    \label{eq:Smatrix_unitary}
    S_{71}S_{71}^\dagger + S_{73} S_{73}^\dagger = I.
\end{equation}
We can reorganize Eq. \ref{eq:Smatrix_unitary} into $I-S_{73} S_{73}^\dagger = S_{71} S_{71}^\dagger $, where the left-hand side is related to the outcoupling, and the right-hand side is related to the diffraction efficiency. In the simplified case of a polarization insensitive grating, these $2\times2$ matrices $S_{71}$ and $S_{73}$ become complex numbers $s_{71}$ and $s_{73}$. Then, the physical meanings of Eq. \ref{eq:Smatrix_unitary} can be straightforwardly interpreted as the following. The diffraction efficiency is $\alpha =|s_{71} |^2$. When the guided light interacts with the input grating, only $|s_{73} |^2=1 - \alpha $ of the guided light remains. Therefore, Eq. \ref{eq:Smatrix_unitary} is a quantitative statement that the higher the diffraction efficiency of the grating is, the larger the chance of outcoupling gets.

We proceed the analysis of the scattering matrix by applying singular value decomposition (SVD) to these $2\times2$ blocks of the scattering matrix \cite{miller2019waves}:
\begin{align}
    \label{eq:S71_SVD} S_{71} & = U_1 \Sigma_1 V_1^\dagger, \\
    \label{eq:S73_SVD} S_{73} & = U_3 \Sigma_3 V_3^\dagger,
\end{align}
where $U_1$ and $U_3$ are the left-singular vectors of $S_{71}$ and $S_{73}$ respectively, $V_1$ and $V_3$ are the right-singular vectors of $S_{71}$ and $S_{73}$ respectively, and $\Sigma_1$ and $\Sigma_3$ are diagonal matrices denoting the singular values of $S_{71}$ and $S_{73}$ respectively. We denote $\Sigma_1$ and $\Sigma_3$ as
\begin{align}
    \label{eq:Sigma1} \Sigma_1 & = \begin{bmatrix} \sqrt{\alpha_1} & 0 \\ 0 & \sqrt{\alpha_2} \end{bmatrix}, \\
    \label{eq:Sigma2} \Sigma_3 & = \begin{bmatrix} \sqrt{1-\alpha_1} & 0 \\ 0 & \sqrt{1 - \alpha_2} \end{bmatrix},
\end{align}
where $\alpha_1$ and $\alpha_2$ have the physical meaning of diffraction efficiency. We assume $\alpha_1 \geq \alpha_2$. If the grating is polarization insensitive at the FOV, $\alpha_1 = \alpha_2$. If the grating is polarization sensitive at the FOV, $\alpha_1 \neq \alpha_2$. With the form of $\Sigma_1$ and $\Sigma_3$ and the unitary constraint (Eq. \ref{eq:Smatrix_unitary}), we find that
\begin{equation}
    \label{eq:U1_U3}
    U_1 = U_3.
\end{equation}

\begin{figure}[tbp]
    \centering
    \includegraphics[width=10cm]{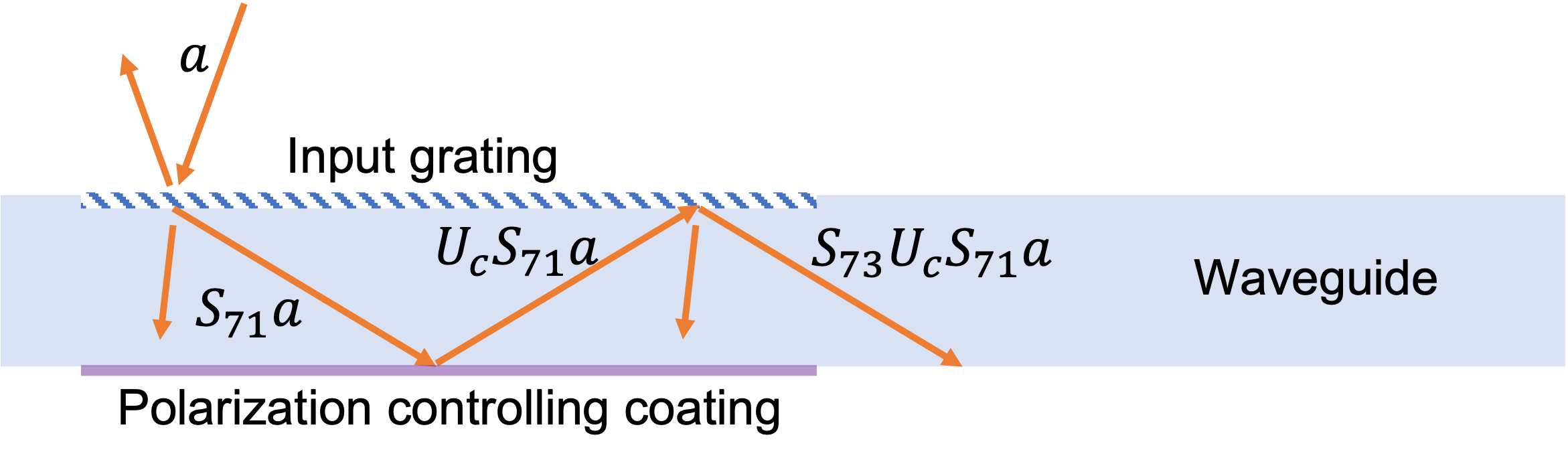}
    \caption{Illustration of the amplitude of incident and guided light as it interacts with the input grating and the polarization controlling coating.}
    \label{fig:illustration_input}
\end{figure}

Without loss of generality, we assume there is a polarization controlling coating in the input grating region, as shown in Fig. \ref{fig:illustration_input}. The polarization management functionality may also be combined into the input grating. To describe the light propagating from the completion of an interaction with the input grating to the onset of the next interaction with the input grating, we use $U_c$, a 2x2 matrix, to represent the round-trip phase accumulation and the reflection phases of the polarization control coating.

\subsection{Uniform input gratings}
We first analyze the input efficiency of a uniform input grating.
With the aforementioned notations, we study the guided power after the light interacts with the input grating for $n$ times. In the first interaction, incident light is diffracted into the guided light, while in the subsequent interactions, part of the guided light may be outcoupled and the rest remains guided. We denote the amplitude of the incident light as $\boldsymbol{a}$, which is a 2-vector representing the amplitudes of two polarizations. To simplify the normalization with respect to the incident power, we assume $\boldsymbol{a}$ is normalized, i.e., $\boldsymbol{a}^\dagger \boldsymbol{a}=1$. After interacting with the grating for the first time, the guided amplitude is $S_{71} \boldsymbol{a}$. With a round-trip traveling in the waveguide and interacting with the grating for the second time, the guided amplitude is $S_{73} U_c S_{71} \boldsymbol{a}$. Thus, after interacting with the grating for $n$ times, the guided amplitude ($\boldsymbol{b}_n$) is 
\begin{equation}
    \label{eq:bn_1}
    \boldsymbol{b}_n = (S_{73} U_c)^{n-1}S_{71} \boldsymbol{a}.
\end{equation}
The corresponding normalized guided power after $n$-time interaction is 
\begin{equation}
    \label{eq:Pn_1}
    P_n = \boldsymbol{b}_n^\dagger \boldsymbol{b}_n.
\end{equation}
With SVD (Eqs. \ref{eq:S71_SVD} and \ref{eq:S73_SVD}) and the condition in Eq. \ref{eq:U1_U3}, we can rewrite Eq. \ref{eq:bn_1} as:
\begin{equation}
    \label{eq:bn_2}
    \boldsymbol{b}_n = U_3 (\Sigma_3 U)^{n-1} \Sigma_1 V_1^\dagger \boldsymbol{a},
\end{equation}
where $U = V_3^\dagger U_c U_3$ and $U$ is a unitary matrix. Similarly, we can rewrite the guided power (Eq. \ref{eq:Pn_1}) as
\begin{equation}
    \label{eq:Pn_2}
    P_n = \boldsymbol{a}^\dagger V_1 \Sigma_1 (U^\dagger \Sigma_3)^{n-1} (\Sigma_3 U)^{n-1} \Sigma_1 V_1^\dagger \boldsymbol{a}.
\end{equation}
When the incident light does not have a pure polarization, the incident polarization ensemble can be represented by a density matrix $\rho$. For instance, for polarized incident light with amplitude $\boldsymbol{a}$, $\rho = \boldsymbol{a} \boldsymbol{a}^\dagger$. For unpolarized incident light, $\rho = \begin{bmatrix} 0.5 & 0 \\ 0 & 0.5 \end{bmatrix}$. For general incident polarization ensemble, the guided power after $n$-time interaction is
\begin{equation}
    \label{eq:Pn_general_polarization}
    P_n = Tr[\rho V_1 \Sigma_1 (U^\dagger \Sigma_3)^{n-1} (\Sigma_3 U)^{n-1} \Sigma_1 V_1^\dagger],
\end{equation}
where $Tr$ represents the trace of a matrix.

Equation \ref{eq:Pn_general_polarization} provides a quantitative calculation of the guided power after interacting with the input grating for $n$ times, with arbitrary incident polarization state ensemble. From Eq. \ref{eq:Pn_general_polarization}, we find that the most significant characteristics of the input grating are the singular values ($\Sigma_1$) of the diffraction Jones matrix $S_{71}$, and its right-singular vectors ($V_1$). The essence of polarization control is to tune $U$ and $V_1$ to control the guided power. $U$ can be controlled with a polarization controlling layer after light first interacts with the input grating or as a part of the input grating, while $V_1$ can be controlled with a polarization controlling layer before the light incident on the input grating or as a part of the input grating. To obtain the upper bound of the input efficiency, we can tune these characteristics of the input grating and polarization controlling layer.

We will show later that the upper bound of the input efficiency of polarization sensitive gratings is higher than that of polarization insensitive gratings. Such improvement exists for either polarized input or unpolarized input. Since the input grating is polarization sensitive, after the light interacts with the input grating for the first time, the diffracted light has dominantly one polarization. With polarization management, the dominant polarization of the diffracted light is converted to the polarization that has the smaller diffraction efficiency when interacting with the input grating again. 

\begin{table}[tbp]
\begin{center}
\begin{tabular}{ | m{2cm} | m{2cm}| m{3cm} | m{3cm} | }
\hline
\multicolumn{2}{|c|}{max $P_2$} & \multicolumn{2}{ c| }{Incident light} \\ \cline{3-4}
\multicolumn{2}{|c|}{Conditions} & Polarized & Unpolarized \\ \cline{1-4}
& Polarization sensitive & 100\% & 50\% \\
& & $\alpha_1=1$, $\alpha_2=0$, & $\alpha_1=1$, $\alpha_2=0$, \\
& & $V_1^\dagger \boldsymbol{a} = [1, 0]^T$, & \\
Input grating & & $U=\begin{bmatrix} 0 & e^{-i\phi_n} \\ e^{i\phi_n} & 0 \end{bmatrix}$ & $U=\begin{bmatrix} 0 & e^{-i\phi_n} \\ e^{i\phi_n} & 0 \end{bmatrix}$  \\ \cline{2-4} 
& Polarization insensitive & 25\% & 25\% \\
& & $\alpha_1=\alpha_2=0.5$ & $\alpha_1=\alpha_2=0.5$ \\
\hline
\end{tabular}
\caption{\label{tab:P2} The maximal guided power after interacting with the input grating twice ($P_2$) in different cases. The conditions to achieve maximal $P_2$ are also shown in the table, where $\phi_n$ is an arbitrary phase, and global phases are omitted.}
\end{center}
\end{table}

As a simplified demonstration, we showcase the maximal $P_2$, which is the guided power after interacting with the input grating for 2 times. Depending on whether the input grating is polarization sensitive or not and whether the incident light is polarized or unpolarized, we show the maximal $P_2$ and the required conditions to achieve the maximal $P_2$ for the 4 cases in Table \ref{tab:P2}. When the input grating is polarization sensitive, the maximal $P_2$ is achieved when $\alpha_1 = 1$ and $\alpha_2 = 0$. The physical meaning is that the input grating diffracts one polarization with 100\% efficiency, while it does not respond to the orthogonal polarization. Such high polarization selectivity has been demonstrated in polarization volume gratings \cite{lee2017reflective, xiong2021holographic, xiang2017nanoscale, weng2016polarization, kobashi2016planar, gao2017high}. Further, the requirement for the polarization management component is $U=\begin{bmatrix} 0 & e^{-i\phi_n} \\ e^{i\phi_n} & 0 \end{bmatrix}=\cos\phi_n\sigma_x + \sin\phi_n\sigma_y$, where $\phi_n$ is an arbitrary phase and the global phase is omitted. The functionality of the polarization controlling layer is to convert the polarization state that is the left-singular vector of $S_{71}$ (or $S_{73}$) corresponding to the larger singular value ($\sqrt{\alpha_1}$) to the polarization state that is the right-singular vector of $S_{73}$ corresponding to the larger singular value ($\sqrt{1-\alpha_2}$). If the incident light is polarized, the maximal $P_2$ is 100\%, which is achieved when the right-singular vector of $S_{71}$ corresponding to the larger singular value ($\sqrt{\alpha_1}$) matches with the incident polarization state ($\boldsymbol{a}$). If the incident light is unpolarized, the maximal $P_2$ is 50\%, which means one polarization gets completely diffracted when it interacts with the input grating for the first time and undergoes no outcoupling when it interacts with the input grating for the second time. On the contrary, when the input grating is polarization insensitive, the maximal $P_2$ is 25\% independent of the incident polarization state ensemble. The maximal $P_2$ is achieved when the diffraction efficiency of the input grating is 50\%.

\begin{figure}[tbp]
    \centering
    \includegraphics[width=8cm]{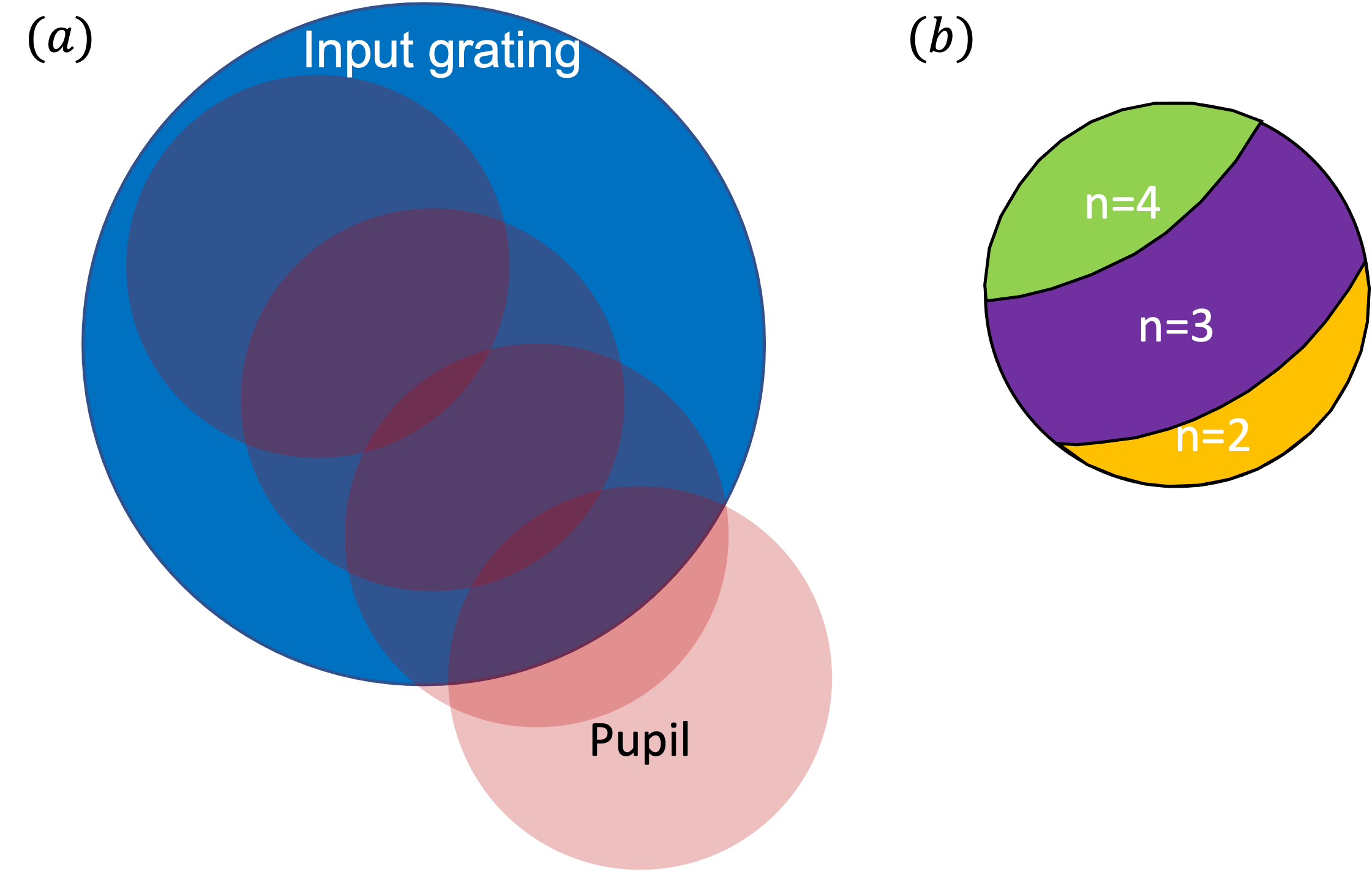}
    \caption{(a) Illustration of the projected incident pupil on the input grating at the first interaction and the subsequent interactions. (b) Segmentation of the incident pupil with respect to $n$, which is the number of interactions with the input grating.}
    \label{fig:pupil_segmentation}
\end{figure}

Next, we segment the incident pupil based on the times of interaction with the input grating for each FOV. This segmentation depends on the geometric and diffraction parameters, such as the projector pupil size and location, the chosen FOV, the wavelength, the input grating size and shape, location, pitch and orientation, and the waveguide index and thickness. Figure \ref{fig:pupil_segmentation}(a) sketches the locations of the incident pupil when it interacts with the input grating, possibly with multiple interactions. Then, we can segment the incident pupil based on the number of interactions with the input grating, as illustrated in Fig. \ref{fig:pupil_segmentation}(b).

With the pupil segmentation and the normalized guided power after $n$ interactions $P_n$, we can calculate the input efficiency by taking the average over different segmentations. If the intensity distribution is uniform, the input efficiency is 
\begin{equation}
    \label{eq:eta_uniform}
    \eta = \frac{\sum_n A_n P_n}{\sum_n A_n},
\end{equation}
where $A_n$ is the area of the incident pupil segmentation with n interactions with the input grating. More generally, if the intensity distribution of the incident light is described by $I(\vec{r})$, the input efficiency is 
\begin{equation}
    \label{eq:eta_varying_intensity}
    \eta = \frac{\sum_n P_n \int_n I(\vec{r}) d^2 \vec{r}}{\int I(\vec{r}) d^2\vec{r}},
\end{equation}
where $\int_n$ means the integration within the pupil segmentation of $n$ interactions. In summary, with Eqs. \ref{eq:Pn_general_polarization} and \ref{eq:eta_varying_intensity}, we can calculate the input efficiency for each FOV deterministically. Formally, the input efficiency ($\eta$) is a function of the incident polarization state ensemble ($\rho$), the scattering matrix blocks of the input grating ($S_{71}$, $S_{73}$) and the polarization control layer ($U_c$), the intensity distribution of the incident light ($I(\vec{r})$), and the set of parameters leading to the segmentation of the incident pupil (denoted as $X$), i.e.,
\begin{equation}
    \label{eq:eta_formal}
    \eta = \eta(\rho, S_{71}, S_{73}, U_c, I(\vec{r}), X).
\end{equation}
The optimization of the input efficiency can be written in the following mathematical form: 
\begin{equation}
\label{eq:eta_maximize}
\begin{aligned}
& \underset{S_{71}, S_{73}, U_c}{\text{maximize}}
& & \eta(\rho, S_{71}, S_{73}, U_c, I(\vec{r}), X) 
\end{aligned}
\end{equation}
Furthermore, with SVD, the only tunable parameters are the diffraction efficiencies $\alpha_1$, $\alpha_2$, and the unitary matrices $U$ (and possibly $V_1$ when the incident light is polarized). The number of independent degrees of freedom in this optimization problem is only a few and we can obtain the global optimum. The optimized $\eta$ is the maximal input efficiency for the FoV, and the corresponding $S_{71}$, $S_{73}$, $U_c$ are the characteristics of the grating and the polarization control layer. In summary, this theoretical model can deterministically calculate and optimize the input efficiency for a uniform input grating.

% ============= Subsection about spatially varying input gratings ====
\subsection{Spatially varying input gratings}

In this section, we extend our discussion to spatially varying input gratings. Compared with a uniform input grating, the only difference is that the characteristics of the input grating ($S_{71}$, $S_{73}$) and the polarization controlling layer ($U_c$) are now spatially dependent. To prepare for this discussion, we introduce a translation operator $T$ to describe the translation of the pupil location from one interaction to the next interaction with the input grating, as illustrated in Fig. \ref{fig:pupil_segmentation}(a). Suppose the displacement between consecutive interactions is $\vec{d} = [d_x, d_y]^T$. The operator describing such translational transformation is 
\begin{equation}
    \label{eq:Translation}
    T_{\vec{d}}\vec{r} = \vec{r} + \vec{d}.
\end{equation}
With $k$ translations, the transformation is denoted as
\begin{equation}
    \label{eq:Translation_k}
    T^k_{\vec{d}}\vec{r} = \vec{r} + k\vec{d}.
\end{equation}

Now, consider the incident light with amplitude $\boldsymbol{a}(\vec{r})$ that is incident at location $\vec{r}$. We define the physical meaning of the amplitude $\boldsymbol{a}(\vec{r})$ such that its modulus square represents the incident intensity at $\vec{r}$. After interacting with the grating for $n$ times, the amplitude of the guided light is
\begin{equation}
    \label{eq:bn_spatial_varying}
    \boldsymbol{b}_n(\vec{r}) = \mathcal{T}\prod_{k=1}^{n-1}\Big[S_{73}\big(T_{\vec{d}}^k \vec{r}\big) U_c\big(T_{\vec{d}}^k \vec{r}\big)\Big] S_{71}(\vec{r}) \boldsymbol{a}(\vec{r}).
\end{equation}
Here, $\mathcal{T}$ represents the ``time-ordered'' product such that larger $k$ terms are on the left side of smaller $k$ terms. For general incident polarization state ensemble described by a density matrix $\rho$, the guided intensity after $n$-time interaction is
\begin{equation}
    \label{eq:In_spatial_varying} 
    I_n(\vec{r}) = Tr\Bigg( \rho(\vec{r}) \Big\{ \mathcal{T}\prod_{k=1}^{n-1}\Big[S_{73}\big(T_{\vec{d}}^k \vec{r}\big) U_c\big(T_{\vec{d}}^k \vec{r}\big)\Big] S_{71}(\vec{r}) \Big\}^\dagger \Big\{ \mathcal{T}\prod_{k=1}^{n-1}\Big[S_{73}\big(T_{\vec{d}}^k \vec{r}\big) U_c\big(T_{\vec{d}}^k \vec{r}\big)\Big] S_{71}(\vec{r}) \Big\} \Bigg) I_\textrm{inc}(\vec{r}).
\end{equation}
Equation \ref{eq:In_spatial_varying} is applicable to spatially varying intensity and spatially varying polarization states. The input efficiency for the spatially varying input grating is
\begin{equation}
    \label{eq:eta_spatial_varying_grating}
    \eta = \frac{\sum_n \int_n I_n(\vec{r}) d^2\vec{r}}{\int I_\textrm{inc}(\vec{r}) d^2\vec{r}}.
\end{equation}
Equation \ref{eq:eta_spatial_varying_grating} is an extension of Eq. \ref{eq:eta_varying_intensity}. With a spatially varying input grating, the number of tunable degrees of freedom is much larger than that of a uniform input grating. Thus, the upper bound of the input efficiency is generally higher than that of a uniform input grating. Nevertheless, it is hard to find the global optimal input efficiency with such a large number of parameters, numerically. Practically, the efficiency limit depends on the acceptable spatial-varying rate, since drastic spatial variations may reduce the resolution of the waveguide combiner. Therefore, in the next section about numerical demonstrations, we focus on uniform gratings with uniform incident intensity distribution.

% =========== Numerical demonstrations ====================
\section{Numerical demonstrations}

In this section, we numerically study a representative example to illustrate our theoretical analysis of the input efficiency. We demonstrate 3 representative cases: (1) The input grating is polarization insensitive. (2) The input grating can be polarization sensitive, and the input light is unpolarized. (3) The input grating is polarization sensitive, and the input light is polarized. To numerically calculate the normalized guided power after interacting with the uniform input grating for $n$ times, we simplify Eq. \ref{eq:Pn_general_polarization} in each case. For each case, we numerically optimize the input grating efficiency using L-BFGS-B method \cite{byrd1995limited, zhu1997algorithm} in the SciPy package \cite{2020SciPy-NMeth}. We present the optimal input grating efficiency and the corresponding grating diffraction efficiency. To characterize the input efficiency and uniformity over the considered FOVs, we calculate the harmonic mean input efficiency over the FOVs and define the min-to-max uniformity as the ratio between the minimal and maximal input efficiency over the FOVs.  

\begin{figure}[tbp]
    \centering
    \includegraphics[width=11cm]{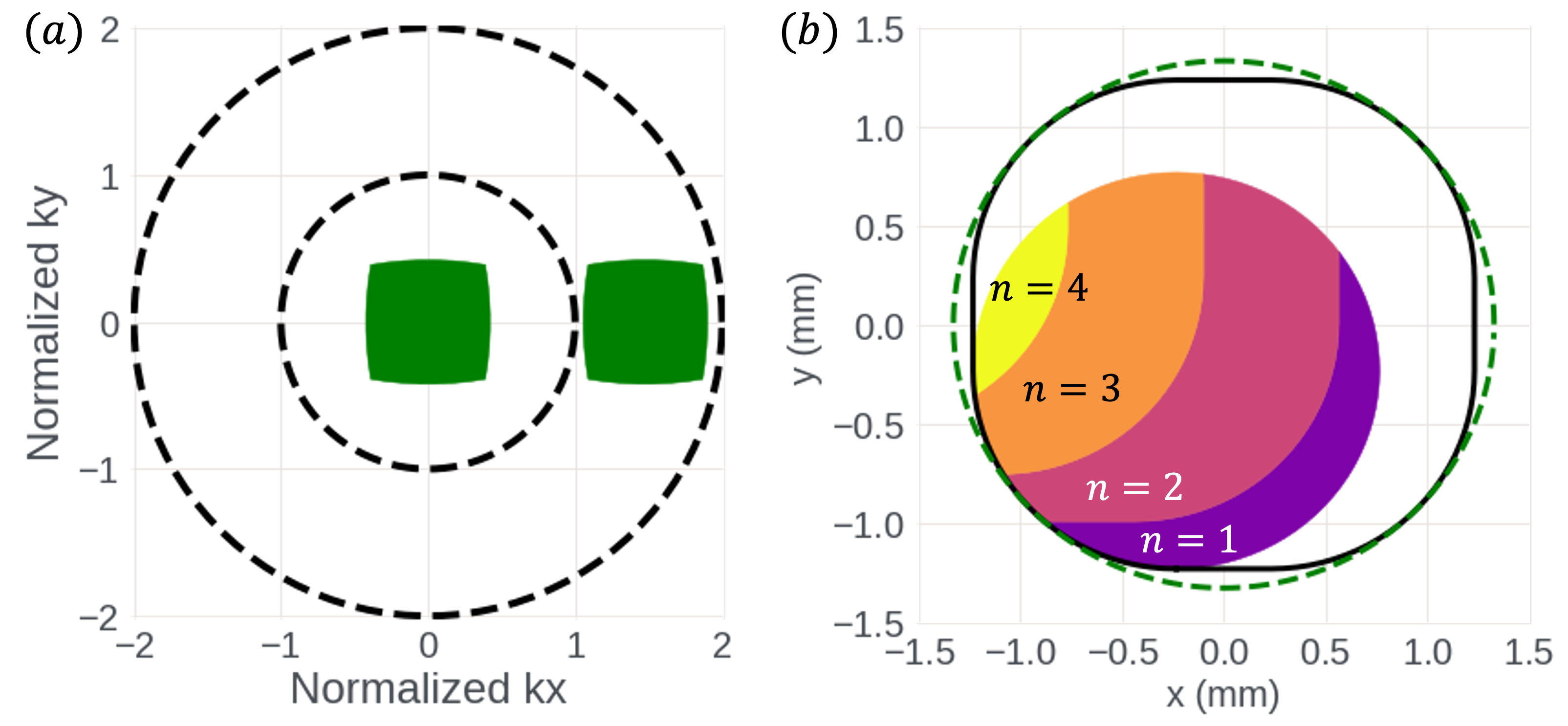}
    \caption{(a) Normalized k-space of $50\times50$ FOV at green wavelength 532 nm, which is diffracted by a grating with pitch 360 nm and wavevector pointing in the x-direction. The transverse wavevectors are normalized with respect to the free space wavevector. We assume the substrate index is 2. The dashed inner circle with radius 1, indicates the largest normalized wavevector in free space. The dashed outer circle, with radius 2, indicates the largest normalized wavevector of propagating waves in the substrate. (b) The black curve represents the minimal covering input grating outline. The dashed green curve represents a circle with radius 1.33 mm. The filled circle is the exit pupil of the projector that is projected to the input grating plane with respect to a FOV ($\theta_x=25^\circ$, $\theta_y=25^\circ$). The different colors represent the pupil segmentation. The purple, pink, orange and yellow regions represent interaction times 1, 2, 3, and 4, respectively.}
    \label{fig:numerical_demo}
\end{figure}

The quantitative numerical results depend on the system parameters. As a representative example, we choose a typical set of system parameters as following: The diameter of a circular exit pupil of the projector is 2 mm. The intensity distribution is uniform over the circular projector pupil. The projector pupil relief distance between the projector exit pupil and the input grating on the waveguide is 0.5 mm. The FOV of the projector is $50\times50$ in degree. The central wavelength of the single-color projector is 532 nm. Our analysis is at this central wavelength. The waveguide is made of glass with index 2. The waveguide thickness is 0.5 mm. To guide the whole FOV, we choose the input grating to have a pitch of 360 nm and a wavevector orientation along the x-direction, as illustrated in Fig. \ref{fig:numerical_demo}(a). We assume that the input grating covers and only covers the whole region that is illuminated by the projector, as illustrated in Fig. \ref{fig:numerical_demo}(b). We refer to such an input grating as the minimal covering input grating. With the aforementioned set of parameters, the minimal covering input grating is slightly smaller than a disk with radius 1.33 mm (Fig. \ref{fig:numerical_demo}(b)). In Fig. \ref{fig:numerical_demo}(b), we also show an example of the pupil segmentation for a specific FOV, where regions represented by different colors have different interaction times with the input grating. The FOV is characterized by two angles $\theta_x$ and $\theta_y$ in the coordinate of a viewer facing the waveguide normally. For instance, the ray normal to the waveguide (and the eyebox) corresponds to a FOV $\theta_x=0$ and $\theta_y=0$. The ray entering the eyebox from the left has a FOV $\theta_x < 0$.

\subsection{Case (1): polarization insensitive input gratings}

When the grating is polarization insensitive, i.e., $\alpha_1=\alpha_2$, Eq. \ref{eq:Pn_general_polarization} can be simplified to 
\begin{equation}
    \label{eq:Pn_case1}
    P_n = \alpha_1 (1 - \alpha_1)^{n-1}.
\end{equation}
In this case, there is only one tunable parameter ($\alpha_1$) in the optimization problem (Eq. \ref{eq:eta_maximize}). 

\begin{figure}[tbp]
    \centering
    \includegraphics[width=10cm]{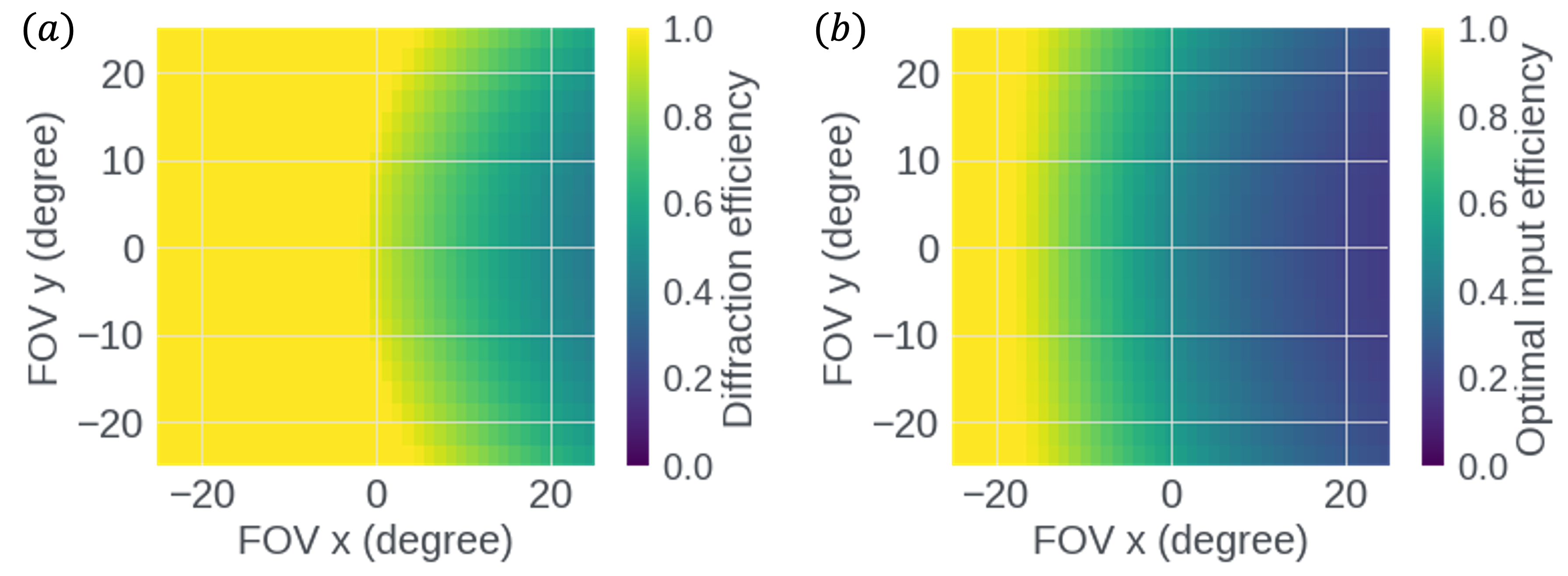}
    \caption{For case (1) where the input grating is polarization insensitive: (a) The optimal grating diffraction efficiency. (b) The optimal input efficiency. }
    \label{fig:efficiency_case1}
\end{figure}

The optimal scalar diffraction efficiency of the input grating over the FOV is shown in Fig. \ref{fig:efficiency_case1}(a). The resulting optimal input efficiency is shown in Fig. \ref{fig:efficiency_case1}(b). For the left-most FOVs, the diffraction angle inside the waveguide is largest and the number of interactions with the input grating is fewest. Thus, the efficiency loss due to the outcoupling is lowest. This is consistent with the numerical results where the left-most FOVs have the highest optimal input efficiency and the right-most FOVs have the lowest optimal input efficiency. To characterize the effective input efficiency across the whole FOV, we take the harmonic mean of the input efficiency over the considered FOVs. In this case, the harmonic mean input efficiency is 42.8\%. We define the ratio between the lowest input efficiency and the highest input efficiency across the FOV as the min-to-max uniformity. In this numerical example, the min-to-max uniformity is 0.18.

\subsection{Case (2): unpolarized incident light}

When the input light is unpolarized, the density matrix $\rho = 0.5I$, where $I$ is a $2\times2$ identity matrix. Equation \ref{eq:Pn_general_polarization} can be simplified to
\begin{equation}
    \label{eq:Pn_case2}
    P_n = 0.5 Tr[\Sigma_1^2(U^\dagger \Sigma_3)^{n-1} (\Sigma_3 U)^{n-1}].
\end{equation}
Since $U$ is a $2\times2$ unitary matrix, it takes the following form:
\begin{equation}
    \label{eq:U_unitary}
    U = e^{i\phi_g}(\cos\theta_u I + i\sin\theta_u \hat{n}\cdot \hat{\sigma}),
\end{equation}
where $I$ is a $2\times2$ identity matrix, $\hat{n} = (n_x, n_y, n_z)$ with $n_x^2 + n_y^2 + n_z^2 = 1$, and $\hat{\sigma} = (\sigma_x, \sigma_y, \sigma_z)$ are Pauli matrices. The global phase $\phi_g$ does not influence the results and we omit it in the following discussions. Furthermore, we denote $n_x = \sin\theta_n \cos\phi_n$, $n_y = \sin\theta_n \sin\phi_n$, and $n_z = \cos\theta_n$. In this case, the input efficiency optimization problem (Eq. \ref{eq:eta_maximize}) has 5 tunable parameters: $\alpha_1$, $\alpha_2$, $\theta_u$, $\theta_n$, and $\phi_n$. 

\begin{figure}[tbp]
    \centering
    \includegraphics[width=14.5cm]{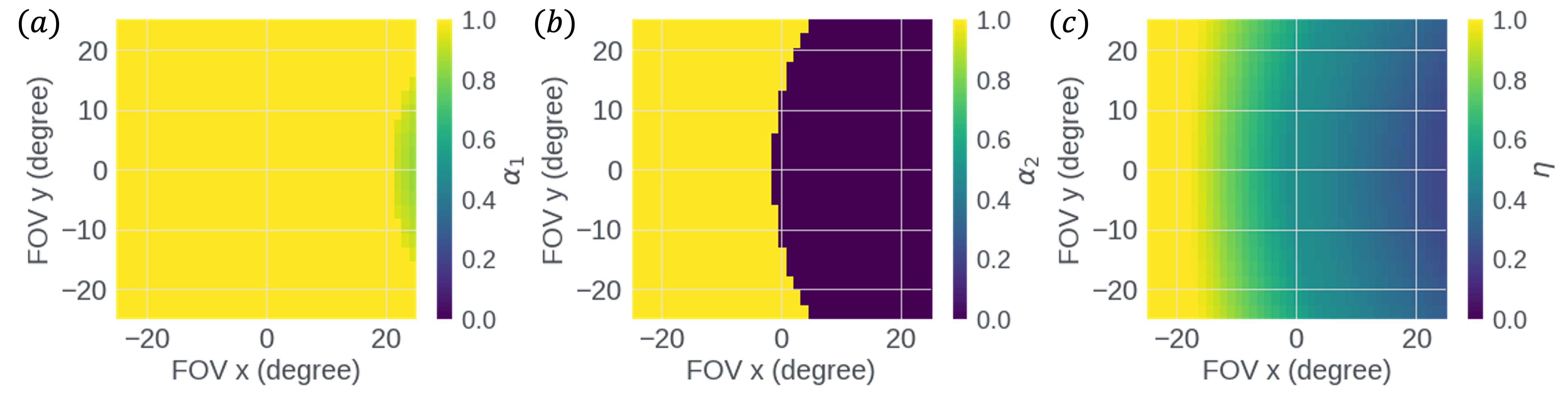}
    \caption{For case (2) where the incident light is unpolarized, the optimal grating diffraction efficiency $\alpha_1$ and $\alpha_2$ are in (a) and (b) respectively. (c) The optimal input efficiency $\eta$. }
    \label{fig:efficiency_case2}
\end{figure}

In this case, the optimal diffraction efficiencies $\alpha_1$ and $\alpha_2$ are shown in Figs. \ref{fig:efficiency_case2}(a) and \ref{fig:efficiency_case2}(b) respectively. The resulting optimal input grating efficiency ($\eta$) is shown in Fig. \ref{fig:efficiency_case2}(c). From the plots of $\alpha_1$ and $\alpha_2$, we find that around the left FOVs, both $\alpha_1$ and $\alpha_2$ are 1; around the middle right FOVs, $\alpha_1$ is 1 and $\alpha_2$ is 0; towards the right-most FOVs, $\alpha_1$ is smaller than 1 and $\alpha_2$ is 0. When both $\alpha_1$ and $\alpha_2$ are 1, the polarization control layer characterized by $U$ is insignificant. When $\alpha_1$ is 1 and $\alpha_2$ is zero, $U = \cos\phi_n \sigma_x + \sin\phi_n \sigma_y$. Here $\phi_n$ can take arbitrary value and $U$ is not unique. When $\alpha_1$ is smaller than 1 and $\alpha_2$ is zero, $U$ is close to $\cos\phi_n \sigma_x + \sin\phi_n \sigma_y$. This is consistent with the previous discussion about $P_2$. The harmonic mean input efficiency is 49.0\%, which is higher than case (1). The min-to-max uniformity is 0.21, which is better than case (1). This suggests that, if the input grating can be polarization selective over certain FOVs, the input grating efficiency and uniformity can be both improved for unpolarized input. This result indicates that it is incorrect to assume that the polarization selective input gratings always have lower input efficiency compared with polarization unselective input gratings when the input light is unpolarized. On the contrary, over certain FOVs, the theoretical upper bound of the input efficiency is higher for the polarization-selective input gratings when the input light is unpolarized. The underlying physics is that though at least half of the input light cannot interact with the polarization selective input grating, the outcoupling can be reduced with polarization management, which is more important for certain FOVs.

\subsection{Case (3): Polarized incident light and polarization sensitive input gratings}

When the input light is polarized, we can denote the input vector as $V_1 \boldsymbol{a}$, such that the density matrix $\rho = V_1 \boldsymbol{a} \boldsymbol{a}^\dagger V_1^\dagger$. Equation \ref{eq:Pn_general_polarization} can be simplified to 
\begin{equation}
    \label{eq:Pn_case3}
    P_n = \boldsymbol{a}^\dagger \Sigma_1 (U^\dagger \Sigma_3)^{n-1} (\Sigma_3 U)^{n-1}\Sigma_1 \boldsymbol{a}.
\end{equation}
For any polarization state denoted by $\boldsymbol{a}$, we can find a unitary matrix $V_1$ such that $V_1\boldsymbol{a}$ matches the incident polarization state. Thus, we can view $\boldsymbol{a}$ as the independent variable in the input efficiency optimization problem (Eq.\ref{eq:eta_maximize}). The general form for $\boldsymbol{a}$ is $\boldsymbol{a} = [\cos\theta_a, \sin\theta_a e^{i\phi_a}]^T$, omitting the global phase. In this case, the input efficiency optimization problem (Eq. \ref{eq:eta_maximize}) has 7 tunable parameters: $\alpha_1$, $\alpha_2$, $\theta_u$, $\theta_n$, $\phi_n$, $\theta_a$, and $\phi_a$. 

\begin{figure}[tbp]
    \centering
    \includegraphics[width=14.5cm]{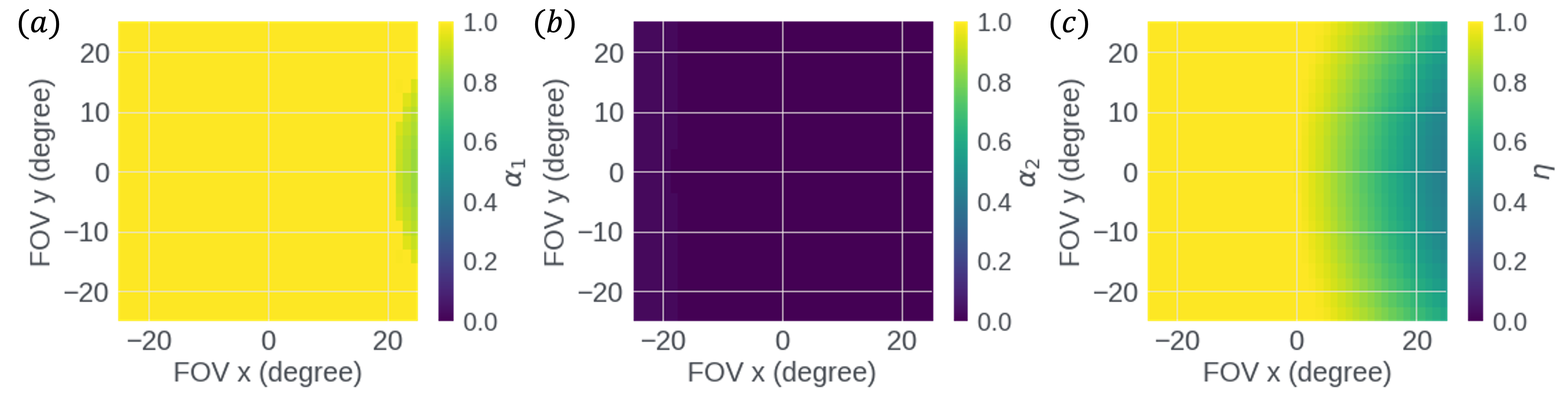}
    \caption{For case (3) where the incident light is polarized, the optimal grating diffraction efficiency $\alpha_1$ and $\alpha_2$ are in (a) and (b) respectively. (c) The optimal input efficiency $\eta$.}
    \label{fig:efficiency_case3}
\end{figure}

For the case of polarized incident light and a polarization sensitive input grating, the optimal diffraction efficiencies $\alpha_1$ and $\alpha_2$ are shown in Figs. \ref{fig:efficiency_case3}(a) and \ref{fig:efficiency_case3}(b) respectively. The resulting optimal input efficiency ($\eta$) is shown in Fig. \ref{fig:efficiency_case3}(c). To achieve the optimal input grating efficiency, the incident light polarization should match the right singular vector of $S_{71}$ with respect to the larger singular value, i.e., $\boldsymbol{a}=[1,0]^T$. $\alpha_1$ is 1 for the most of the FOV and becomes less than 1 towards the right FOV. $\alpha_2$ is zero for the entire FOV. However, at the left FOV when light only interacts with the input grating once for the entire pupil, both $\alpha_2$ and the polarization control layer $U$ can take arbitrary value, since they are not involved in the input grating efficiency calculation. For the rest FOV region, $\alpha_2=0$ deterministically. When $\alpha_1=1$, and $\alpha_2=0$, the polarization control layer $U$ is typically $\cos\phi_n \sigma_x + \sin\phi_n \sigma_y$, where $\phi_n$ is arbitrary. When $\alpha_1<1$, and $\alpha_2=0$, the polarization control layer $U$ is close to $\cos\phi_n \sigma_x + \sin\phi_n \sigma_y$. This is consistent with the previous discussion about $P_2$ and our intuitions. When $U=\begin{bmatrix}0 & e^{-i\phi_n} \\ e^{i\phi_n} & 0 \end{bmatrix}$, and $\boldsymbol{a}=[1,0]^T$, the polarized incident light interacts with the larger diffraction efficiency of the input grating initially. After being diffracted, the guided light interacts with the smaller diffraction efficiency ($\alpha_2$) and then the larger diffraction efficiency ($\alpha_1$) rotationally. Since $\alpha_2=0$, the out-coupling when the guided light interacts with the input grating for the second time is eliminated. Thus, this input efficiency is close to case (1) when the thickness of the waveguide is doubled. In case (3), the harmonic mean input efficiency is 83.3\%, and the min-to-max uniformity is 0.42, which are better than cases (1) and (2). 

\subsection{Comparing the 3 cases}

\begin{figure}[tbp]
    \centering
    \includegraphics[width=14cm]{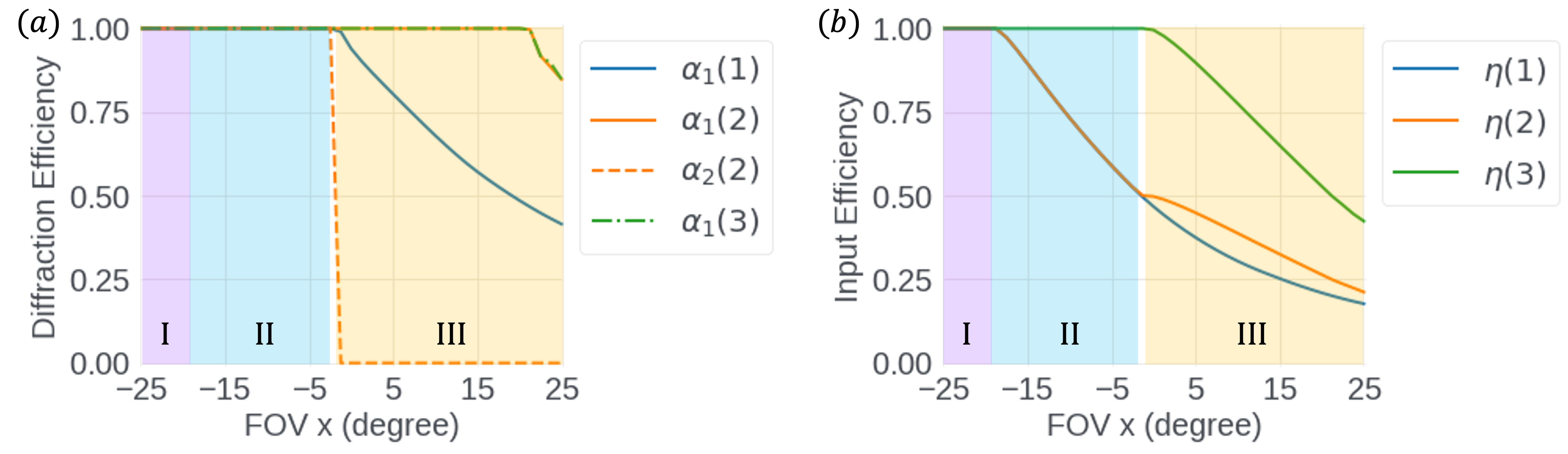}
    \caption{(a) and (b) show $\alpha_1$ (and $\alpha_2$ in case (2)) and $\eta$ in the 3 cases as a function of the horizontal FOV when the vertical FOV is zero degree, respectively. The shaded region corresponds to (I) when the entire pupil only interacts with the input grating once; (II) when part of the pupil interacts with the input grating once and the rest of the pupil interacts with the input grating twice; and (III) when input efficiency in case (2) is higher than that in case (1).}
    \label{fig:comparison}
\end{figure}

To further compare the 3 cases, we plot $\alpha_1$ (and $\alpha_2$ in case (2)) and $\eta$ in the 3 cases as a function of the horizontal FOV when the vertical FOV is zero degree in Figs. \ref{fig:comparison}(a) and \ref{fig:comparison}(b). As the horizontal FOV increases (from left FOV to right FOV), the number of interactions with the input grating increases. Thus, generally speaking, the input grating efficiency decreases, and the optimal grating diffraction efficiency decreases accordingly. We emphasize 3 FOV regions as shaded by purple (I), blue (II) and orange (III) in Figs. \ref{fig:comparison}(a) and \ref{fig:comparison}(b). In FOV region I, the entire pupil only interacts with the input grating once. Thus, there is no out-coupling issue, and the input grating efficiency can reach 100\% for all 3 cases if the diffraction efficiency is 100\%. In FOV region II, part of the pupil interacts with the input grating once and the rest of the pupil interacts with the input grating twice. With polarization management, the input grating efficiency can still reach 100\% if the incident light is purely polarized. In FOV region III, input efficiency in case (2) is higher than that in case (1). This indicates that the efficiency loss due to out-coupling plays a more important role over the efficiency loss due to irresponsiveness to one polarization. This leads to the surprising conclusion that, under certain conditions, polarization management can increase the input efficiency even when the incident light is unpolarized.

We summarize the harmonic mean input efficiency and the min-to-max uniformity for the 3 cases in Table \ref{tab:comparison}. When the incident light is polarized, the input grating with polarization engineering can double the efficiency and uniformity performances compared with the polarization insensitive input grating. Even when the incident is unpolarized, the polarization engineering at the input grating can still improve the efficiency and uniformity by a few percentages. This observation is counter-intuitive without such thorough investigation.

\begin{table}[tbp]
\begin{center}
\begin{tabular}{ | m{4cm} | m{2cm}| m{2cm} | m{2cm} | }
\hline
& Case (1) & Case (2) & Case (3) \\
\hline
Polarization management at the input coupler & No & Yes & Yes \\
\hline
Incident light polarization & Arbitrary & Unpolarized & Polarized \\
\hline
Harmonic mean efficiency & 42.8\% & 49.0\% & 83.3\% \\
\hline
Min-to-max uniformity & 0.18 & 0.21 & 0.42 \\ 
\hline
\end{tabular}
\caption{\label{tab:comparison} Summary of the harmonic mean input efficiency and min-to-max uniformity for cases (1)-(3).}
\end{center}
\end{table}

\section{Discussion}
\label{sec:discussion}

Equipped with our theory, we investigate the efficiency limit of uniform input gratings as a function of waveguide thickness, projector pupil size and projector pupil relief distance in these 3 cases. We use the harmonic mean efficiency over $50\times50$ FOV as the metric to characterize the input efficiency and the min-to-max uniformity as the metric of characterize the input uniformity. When the geometric parameters are varied, the input grating shape changes accordingly such that it is always the minimal covering input grating.

\subsection{Influence of waveguide thickness}

\begin{figure}[tbp]
    \centering
    \includegraphics[width=14cm]{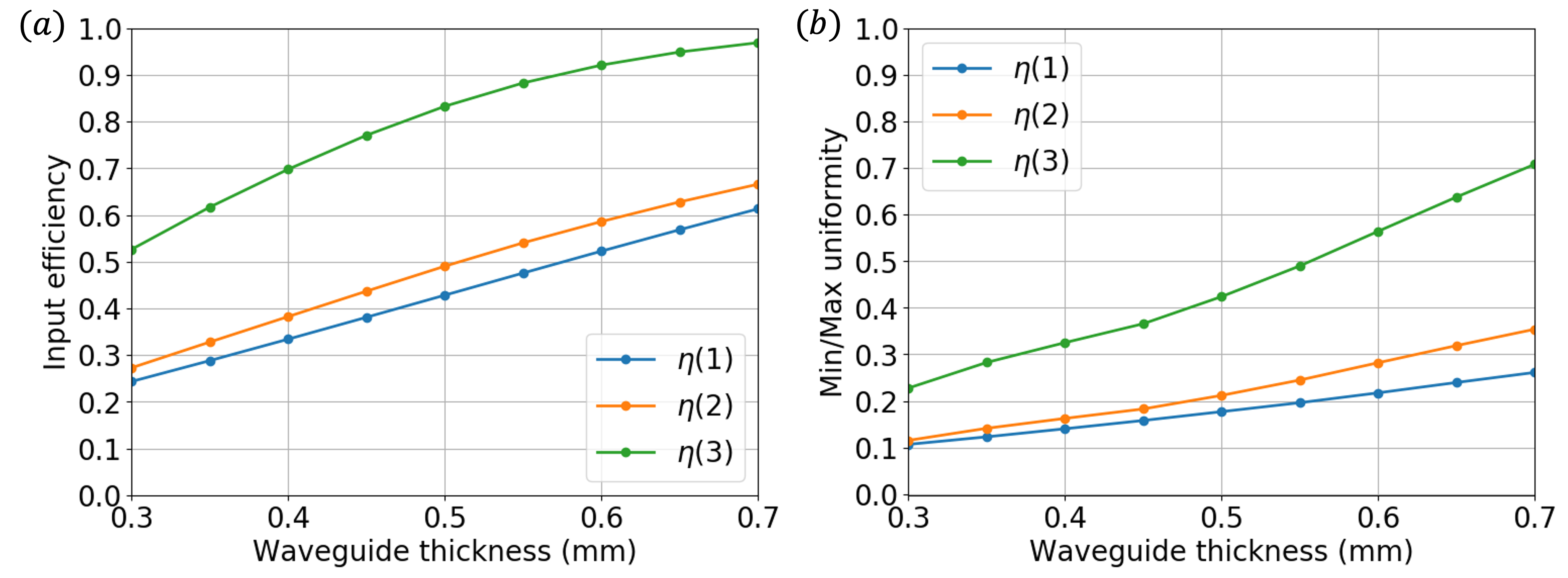}
    \caption{The harmonic mean input efficiency (a) and the min-to-max uniformity (b) as a function of waveguide thickness for the 3 cases. The pupil diameter is 2 mm and the projector pupil relief is 0.5 mm.}
    \label{fig:thickness_sweep}
\end{figure}

We vary the waveguide thickness from 0.3 mm to 0.7 mm while maintaining the values of other parameters. The harmonic mean input efficiency and the min-to-max uniformity as a function of waveguide thickness are shown in Figs. \ref{fig:thickness_sweep}(a) and \ref{fig:thickness_sweep}(b), respectively. The blue, orange, and green curves represent cases (1), (2), and (3) respectively. The input efficiency decreases as the waveguide thickness decreases. Physically, the number of interactions with the input grating increases as the waveguide thickness decreases, and the out-coupling problem becomes more prominent. Thus, besides the mechanical challenges, the low input efficiency is another reason that thin waveguide (e.g., 0.3 mm) is unfavorable. 

\subsection{Influence of projector exit pupil size}

\begin{figure}[tbp]
    \centering
    \includegraphics[width=14cm]{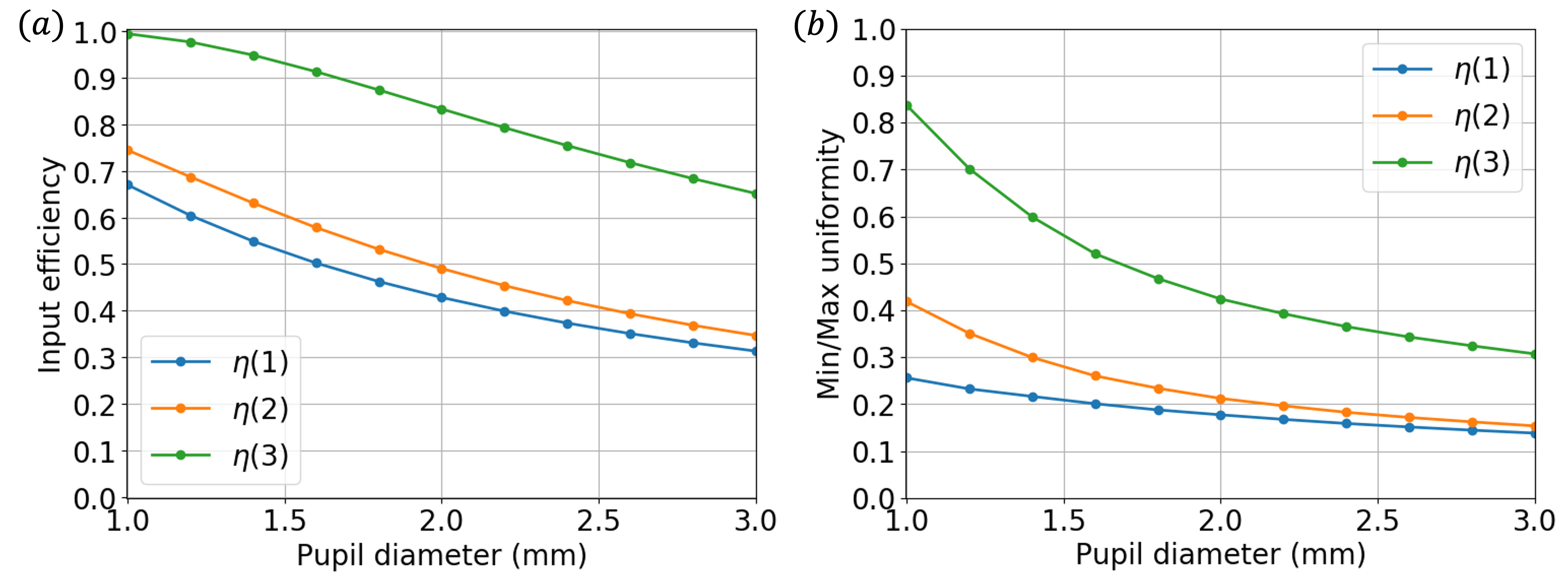}
    \caption{The harmonic mean input efficiency (a) and the min-to-max uniformity (b) as a function of projector pupil diameter for the 3 cases. The waveguide thickness is 0.5 mm and the projector pupil relief is 0.5 mm.}
    \label{fig:pupil_sweep}
\end{figure}

We vary the diameter of the projector exit pupil from 1 mm to 3 mm while maintaining the values of other parameters. The harmonic mean input efficiency and the min-to-max uniformity as a function of projector exit pupil diameter are shown in Figs. \ref{fig:pupil_sweep}(a) and \ref{fig:pupil_sweep}(b), respectively. The blue, orange, and green curves represent cases (1), (2), and (3) respectively. The input efficiency decreases as the projector exit pupil diameter increases. Thus, it is desirable, from the input efficiency point of view, to reduce the projector exit pupil size. Nevertheless, small projector pupil size may cause resolution drop or pupil replication density artifacts \cite{jang2022waveguide}. Therefore, we need to balance these factors in choosing the projector pupil size.

\subsection{Influence of projector pupil relief distance}

\begin{figure}[tbp]
    \centering
    \includegraphics[width=14cm]{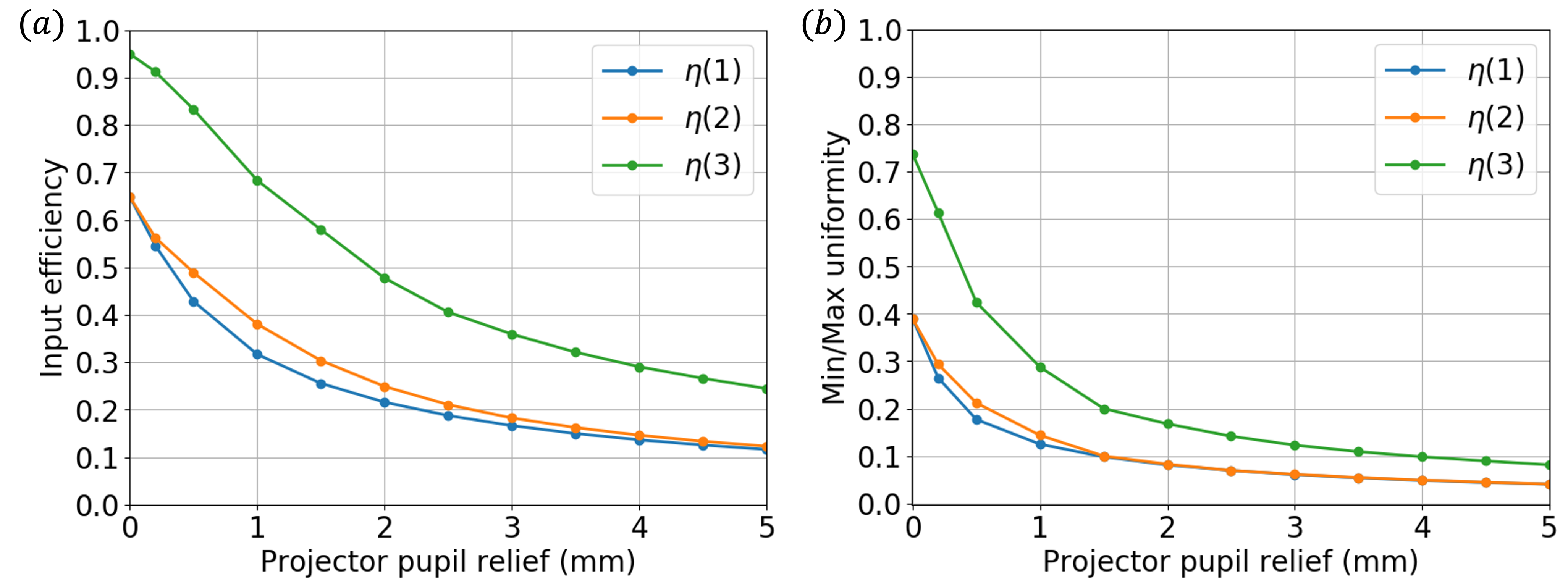}
    \caption{The harmonic mean input efficiency (a) and the min-to-max uniformity (b) as a function of projector pupil relief distance for the 3 cases. The waveguide thickness is 0.5 mm and the pupil diameter is 2 mm.}
    \label{fig:pupil_relief_sweep}
\end{figure}

We also study how the projector pupil relief, which is the distance between the projector exit pupil and the input grating on the waveguide surface, affects the input efficiency. We vary the projector pupil relief distance from 0 mm to 5 mm while keeping the rest parameters unchanged. The harmonic mean input efficiency and the min-to-max uniformity as a function of the pupil relief are shown in Figs. \ref{fig:pupil_relief_sweep}(a) and \ref{fig:pupil_relief_sweep}(b), respectively. The blue, orange, and green curves represent cases (1), (2), and (3) respectively. The input efficiency decreases as the pupil relief distance increases. One challenge of laser beam scanning projectors is the large projector pupil relief distance, which can lead to poor input efficiency. A potential solution to increase the input efficiency is to introduce spatial variation in the input grating. This solution is especially attractive for large pupil relief distance, since the input grating size is also large, and the projected pupils for different FOVs are spatially dispersed. 

\section{Conclusion}
\label{sec:conclusion}

In summary, we provide a systemmatic theoretical investigation of the efficiency limit of diffractive input couplers with general configurations and with the consideration of polarization management. We discuss the physics of the out-coupling problem that constrain the input efficiency and mathematically identify the key parameters that determine the input efficiency. Our theory can deterministically obtain the optimal input efficiency, along with the corresponding characteristics of the input grating and the polarization management component.
Our theory includes both uniform input gratings and spatial varying input gratings and works for polarization sensitive or insensitive input gratings and arbitrary incident polarization state ensemble. We present numerical demonstrations on uniform input gratings and showcase 3 representative cases: (1) polarization insensitive input gratings, (2) unpolarized incident light, and (3) polarized incident light with polarization sensitive input gratings. Our demonstration shows that polarization management can improve the efficiency limit of input gratings. To shine light on the system-level geometric parameters, we apply our theory to study how the optimal input efficiency of a uniform input grating varies with the waveguide thickness, the projector pupil size, and the projector pupil relief distance.
Our theory provides a fundamental understanding of the physics mechanisms governing the input efficiency and highlights the advantages of polarization management. 
Our study is important in the pursuit of AR waveguide combiners with high efficiency that eventually lead to AR glasses with high-quality display, compact form factor and wide accessibility.

%%%%%%%%%%%%%%%%%%%%%%% References %%%%%%%%%%%%%%%%%%%%%%%%%

%%%%%%%%%% If using BibTeX:
\bibliographystyle{unsrt}
\bibliography{wg_bib}
% \printbibliography

%%%%%%%%%% If preparing manually:
% \begin{thebibliography}{1}
% \newcommand{\enquote}[1]{``#1''}

% \bibitem{Zhang:14}
% Y.~Zhang, S.~Qiao, L.~Sun, Q.~W. Shi, W.~Huang, L.~Li, and Z.~Yang,
%   \enquote{Photoinduced active terahertz metamaterials with nanostructured
%   vanadium dioxide film deposited by sol-gel method,}
%   {\protect\JournalTitle{Optics Express}} \textbf{22}, 11070--11078 (2014).

% \bibitem{Optica}
% {Optica}, \enquote{{Optica Publishing Group},}
%   \url{http://www.opg.optica.org}.

% \bibitem{FORSTER2007}
% P.~Forster, V.~Ramaswamy, P.~Artaxo, T.~Bernsten, R.~Betts, D.~Fahey,
%   J.~Haywood, J.~Lean, D.~Lowe, G.~Myhre, J.~Nganga, R.~Prinn, G.~Raga,
%   M.~Schulz, and R.~V. Dorland, \enquote{Changes in atmospheric consituents and
%   in radiative forcing,} in \enquote{Climate Change 2007: The Physical Science
%   Basis. Contribution of Working Group 1 to the Fourth assesment report of
%   Intergovernmental Panel on Climate Change,}  S.~Solomon, D.~Qin, M.~Manning,
%   Z.~Chen, M.~Marquis, K.~B. Averyt, M.~Tignor, and H.~L. Miler, eds.
%   (Cambridge University Press, 2007).

% \end{thebibliography}

\end{document}